\documentclass[twoside,fleqn]{article}
\usepackage{espcrc2}
\usepackage{graphicx,cite}

\newcommand{\figwidth}{8.0cm}
\newcommand{\figheight}{4.5cm}

\newcommand{\nn}{\nonumber}

\newcommand{\mev}{\mbox{\rm MeV}}
\newcommand{\gev}{\mbox{\rm GeV}}
\newcommand{\eqn}[1]{(\ref{#1})}

\newcommand{\ep}{\epsilon}

\newcommand{\pslh}{p\!\!\!\!\!\not\,\,\,}

\newcommand\lsim{\mathrel{\rlap{\lower4pt\hbox{\hskip1pt$\sim$}}
    \raise1pt\hbox{$<$}}}
\newcommand\gsim{\mathrel{\rlap{\lower4pt\hbox{\hskip1pt$\sim$}}
    \raise1pt\hbox{$>$}}}

\title{
\vspace{-1.0cm}
{\sf \small \rightline{IFIC/04-45, FTUV/04-0803}}
\bigskip
{\Large \bf Pentaquark from QCD sum rules: consequences of the diquark
approach}
}
\author{M. Eidem\"uller\thanks{Talk given at the High-Energy Physics
        International Conference in Quantum Chromodynamics
        (QCD 2004), Montpellier, July 2004}
\address{
       {\em Departament de F\'{\i}sica Te\`orica, IFIC,
       Universitat de Val\`encia -- CSIC,}\\
       {\em Apt. Correus 22085, E-46071 Val\`encia, Spain}}
        }

\begin{document}

\begin{abstract}
\noindent
In this work we investigate the consequences of the Jaffe and Wilczek diquark
model in the framework of QCD sum rules. An analysis of the $\Theta^+(1540)$
as  $(ud)^2\bar{s}$ state shows that the mass of the pentaquark is 
compatible with the experimentally measured value.
The mass difference between the $\Theta^+$ and
the pentaquark with the quantum numbers of the nucleon amounts to
70 MeV and is consistent with the interpretation of the $N(1440)$ as a pentaquark.
\end{abstract}

\maketitle

%\vfill

%\noindent
%{\it Keywords}: Pentaquark, QCD sum rules\\
%{\it PACS}: 12.38.Lg, 12.90.+b 

%\end{titlepage}

%\newpage
%\setcounter{page}{1}

%%%%%%%%%%%%%%%%%%%%%%%%%%%%%%%%%%%%%%%%%%%%%%%%%%%%%%%%%%%%%%%%%%%%%%%%%
% Beginning of the paper
%%%%%%%%%%%%%%%%%%%%%%%%%%%%%%%%%%%%%%%%%%%%%%%%%%%%%%%%%%%%%%%%%%%%%%%%%

%%%%%%%%%%%%%%%%%%%  Introduction %%%%%%%%%%%%%%%%%%%%%%%%%%%%%%%%%%%%%%%

\noindent
Recently, several experiments
\cite{Penta_Measurements}
%\cite{Nakano:2003qx,Barmin:2003vv,Stepanyan:2003qr,Barth:2003ja,
%Kubarovsky:2003fi,Airapetian:2003ri,Asratyan:2003cb,
%Aleev:2004sa,:2004kn,Abdel-Bary:2004ts}
have observed a new baryon resonance
$\Theta^+(1540)$ with positive strangeness. Therefore it 
requires an $\bar{s}$ and has a 
minimal quark content of five quarks. 
The $\Theta$ has the third component of isospin zero 
and the absence of isospin partners suggests strongly that the $\Theta$ 
is an isosinglet what we also assume in this work. 
A puzzling characteristics of the $\Theta$ is its narrow width below
15 MeV. A suggestive way to explain the small width is by the assumption of
diquark clustering. The formation of diquarks presents an important concept
and has direct phenomenological impact \cite{Diquarks}.
In this work we investigate the diquark model by Jaffe and Wilczek 
\cite{Jaffe:2003sg,Jaffe:2004zg} in the framework of QCD sum rules where
the $\Theta$ is described as bound state of an $\bar{s}$ with two 
highly correlated $(ud)$-diquarks.
The basis of the sum rules was laid
in \cite{SRbasis} and their extension to baryons was developed in
\cite{SRbaryons}.
The assumptions of the model are
incorporated by an appropriate current. 
Since the sum rules are directly based on QCD and keep the analytic dependence
on the input parameters, they can help to differentiate between the
models and to test their features.
The relevance of the diquark picture within the context of the 
sum rules was shown in \cite{SRdiquarks}.
Several sum rule investigations for the pentaquark already exist 
\cite{Eidemuller:2004ra,Sugiyama:2003zk,Zhu:2003ba,Matheus:2003xr,
Matheus:2004gx,Huang:2003bu,Kondo:2004cr}
which, however, are based on different models or currents.
The diquark models for the pentaquark have also been investigated 
within other approaches \cite{Pentaapproaches}.

In the model by Jaffe and Wilczek
the $(ud)$-diquarks have zero spin and are in a $\bar{3}_c$
and $\bar{3}_f$ representation of colour and flavour. In order to combine with
the antiquark into a colour singlet, the two diquarks must combine into a
colour 3. The diquark-diquark wavefunction is antisymmetric and has angular
momentum one. This combines with the spin of the $\bar{s}$ to total 
angular momentum 1/2 and results in positive parity. 
In \cite{Jaffe:2003sg} it was suggested to interpret the Roper resonance 
$N(1440)$ as $(ud)^2\bar{d}$ pentaquark state and
we will study this resonance at the end of our analysis.

%%%%%%%%%%%%%%% Definitions %%%%%%%%%%%%%%%%%%%%%%%%%%%%%%%%%%%

The basic object in our sum rule analysis is the two-point correlation
function 
\begin{equation}
 \label{eq:PiDef}
\Pi(p) = i \int d^4 x\ e^{ipx} \langle 0|T\{\eta(x)\bar{\eta}(0)\}|0\rangle\,,
\end{equation}
where $\eta(x)$ represents the interpolating field of the 
pentaquark under investigation. 

The diquarks have a particularly strong attraction in the flavour antisymmetric
$J^P=0^+$ channel. Thus the current contains two diquarks of the form
\begin{equation}
 \label{eq:Diquark}
{\cal Q}^c(x) = \ep^{abc}\,Q_{ab}(x) = \ep^{abc}\,[u_a^T C\gamma_5 d_b](x)\,.
\end{equation}
$C$ denotes the charge conjugation matrix.
The two diquarks must be in a $p$-wave to satisfy Bose statistics.
Therefore the current contains a derivative to generate one unit of angular
momentum. The diquarks couple to a $3_c$ in colour to form the current
\begin{eqnarray}
 \label{eq:Current}
\eta(x)&=& \left(\ep^{abd}\delta^{ce}-\ep^{abc}\delta^{de}\right)
[Q_{ab}\left(D^\mu Q_{cd}\right) \nn\\
&& -\left(D^\mu Q_{ab}\right)Q_{cd} ]\gamma_5 \gamma_\mu C \bar{s}_e^T\,,
\end{eqnarray}
where the covariant derivative for the $\bar{3}_c$ is given by
$D^\mu=\partial^\mu-ig \lambda_l^\dagger A^{\mu\,l}$ \cite{Jaffe:2004zg}.
The parity is positive.
This current has a different structure than the currents in other pentaquark
sum rules \cite{Sugiyama:2003zk,Zhu:2003ba,Matheus:2003xr,
Matheus:2004gx,Huang:2003bu,Kondo:2004cr} 
which contain no derivative to produce the angular
momentum between the diquarks. Inserting the current and
neglecting higher orders in the strong coupling constant the correlator is
given by 
\begin{eqnarray}
 \label{eq:Correlator}
\Pi(x)&=& \langle 0|T\{\eta(x)\bar{\eta}(0)\}|0\rangle\nn\\
&=& \left[\gamma_5\gamma^\mu C S^{(s)\,T}_{e'e}(-x) C \gamma^\nu\gamma_5\right] 
T_{\mu\nu}^{ee'}(x)\,, 
 \end{eqnarray}
where $S^{(s)}(x)$ represents the strange quark propagator.
The quark propagator has been evaluated in the presence of quark and gluon
condensates in \cite{Novikov:1985gd,Yang:1993bp,Matheus:2003xr}, 
where the explicit expressions can be found.
Using the following Lorentz decomposition for 
$T_{\mu\nu}^{ee'}=\delta^{ee'} T_{\mu\nu}/3$,
\begin{equation}
 \label{eq:Lorentz}
T_{\mu\nu} = g_{\mu\nu}f_1(x^2)+ x_\mu x_\nu f_2(x^2) \,,
\end{equation}
the functions $f_1(x^2)$ and $f_2(x^2)$ can be determined up to
operators of dimension 6 \cite{Eidemuller:2004ra}.
In momentum space the correlator can be parametrised as
\begin{equation}
 \label{eq:MomCorr}
\Pi(p)=\pslh \Pi^{(p)}(p^2) + \Pi^{(1)}(p^2)\,.
\end{equation}
To obtain the phenomenological side we insert intermediate baryon states with
the corresponding quantum numbers. 
Since no experimental information on higher pentaquark states is available we
make the assumption of quark-hadron duality and approximate the higher states
by the perturbative spectral density above a threshold $s_0$. In fact, the
uncertainty on $s_0$ will be one of the dominant errors in the sum rule
analysis.

In order to suppress the higher dimensional condensates and to reduce the
influence of the higher resonances we employ a Borel
transformation with Borel parameter $M$.
As in \cite{Zhu:2003ba,Matheus:2003xr} we now concentrate on the chirality even part
$\Pi^{(p)}$ in eq. \eqn{eq:MomCorr} which contains the leading order term from
the operator product expansion.
Transferring the continuum contribution to the theoretical side and taking a
logarithmic derivative with respect to $-1/M^2$, one obtains the sum rule for
the mass of the pentaquark,
\begin{eqnarray}
 \label{eq:Mass}
m_\Theta^2 &=& \frac{\sum\limits_{k=0}^{k=3} a_{6-k} \Gamma(8-k)
(M^2)^{8-k} E_{7-k}}
{\sum\limits_{k=0}^{k=3} a_{6-k} \Gamma(7-k) (M^2)^{7-k} E_{6-k}}\,,
\end{eqnarray}
where $E_\alpha=1-\Gamma(\alpha+1,s_0/M^2)/\Gamma(\alpha+1)$.

%%%%%%%%%%%%%%%%%%%%%%%%%% Analysis %%%%%%%%%%%%%%%%%%%%%%%%%%%%%%%%%%%%

A basic input for the sum rule analysis is the Borel parameter $M$.
We employ a sum rule window of
$2.5\ \gev^2 < M^2 < 4.0\ \gev^2$ where the operator product expansion 
converges well and the phenomenological continuum is not too large.
For the continuum threshold we use a central value of 
$s_{0}=(1.54+0.35\ \gev)^2$. Thus the continuum starts 350 MeV above the 
measured pentaquark mass. This difference should roughly correspond to one
radial excitation \cite{Zhu:2003ba} and represents 
a typical value for sum rule analyses with light
quarks as degrees of freedom \cite{SRbasis}.
\begin{figure}
\begin{center}
\includegraphics[height=\figheight,width=\figwidth,angle=0]{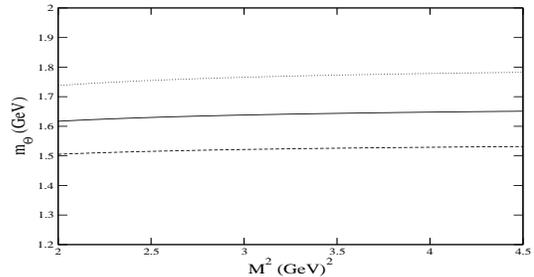}
\caption{\label{fig:1} $m_\Theta$ as a function of the Borel parameter $M^2$
for different $s_0=3.5\ \gev^2$ (solid), $s_0=4.1\ \gev^2$ (dotted) and 
$s_0=3.0\ \gev^2$ (dashed).}
\end{center}
\end{figure}
Fig. \ref{fig:1} shows the mass as a function of the Borel parameter
$M^2$. The sum rule has a good stability with respect to $M$.
As central value for the pentaquark mass we obtain $m_\Theta=1.64\ \gev$.
To estimate the error on $m_\Theta$ we vary $s_0$ between
$3.0\ \gev^2 < s_0 < 4.1\ \gev^2$. 
In fig. \ref{fig:1} we have also plotted the change of $m_\Theta$ with the
continuum threshold from which we obtain an error of 
$\Delta m_\Theta \approx 125\ \mev$.
To estimate the dependence of the sum rules on the OPE we successively remove
the different orders. 
The inclusion of the higher condensates lowers the mass.
The four-dimensional condensates
lower the leading order result by about 50 MeV and the condensates of
dimension 6 by another 50 MeV. We assume that a reasonable error estimate
from the OPE would be $\Delta m_\Theta \approx 75\ \mev$. 
Furthermore, contributions to the error also arise from the other input 
parameters which we vary in the ranges presented above. 
As it turns out, their influence on the value of $m_\Theta$ is
small compared to the errors from the continuum threshold 
and the convergence of the OPE.
Adding the errors quadratically our final result reads
\begin{equation}
 \label{eq:PentaMass}
m_\Theta = 1.64 \pm 0.15 \ \gev.
\end{equation}
In \cite{Jaffe:2003sg} Jaffe and Wilczek suggested to interpret the 
Roper resonance as $(ud)^2\bar{d}$ pentaquark state. 
One can then perform a similar analysis for the $N(1440)$ as has been done
for the $\Theta$ by substituting the $\bar{s}$ antiquark by a 
$\bar{d}$ antiquark. 
As central value for the continuum threshold we choose, as in the $\Theta^+$
case, a value of 350 MeV above the ground state mass. For the error range we
use $2.7\ \gev^2 < s_{0N} < 3.8 \ \gev^2$.
Performing a sum rule analysis for the $N$ with the above given parameters,
we obtain a mass of $m_N=1.57 \pm 0.15\ \gev$.
\begin{figure}
\begin{center}
\includegraphics[height=\figheight,width=\figwidth,angle=0]{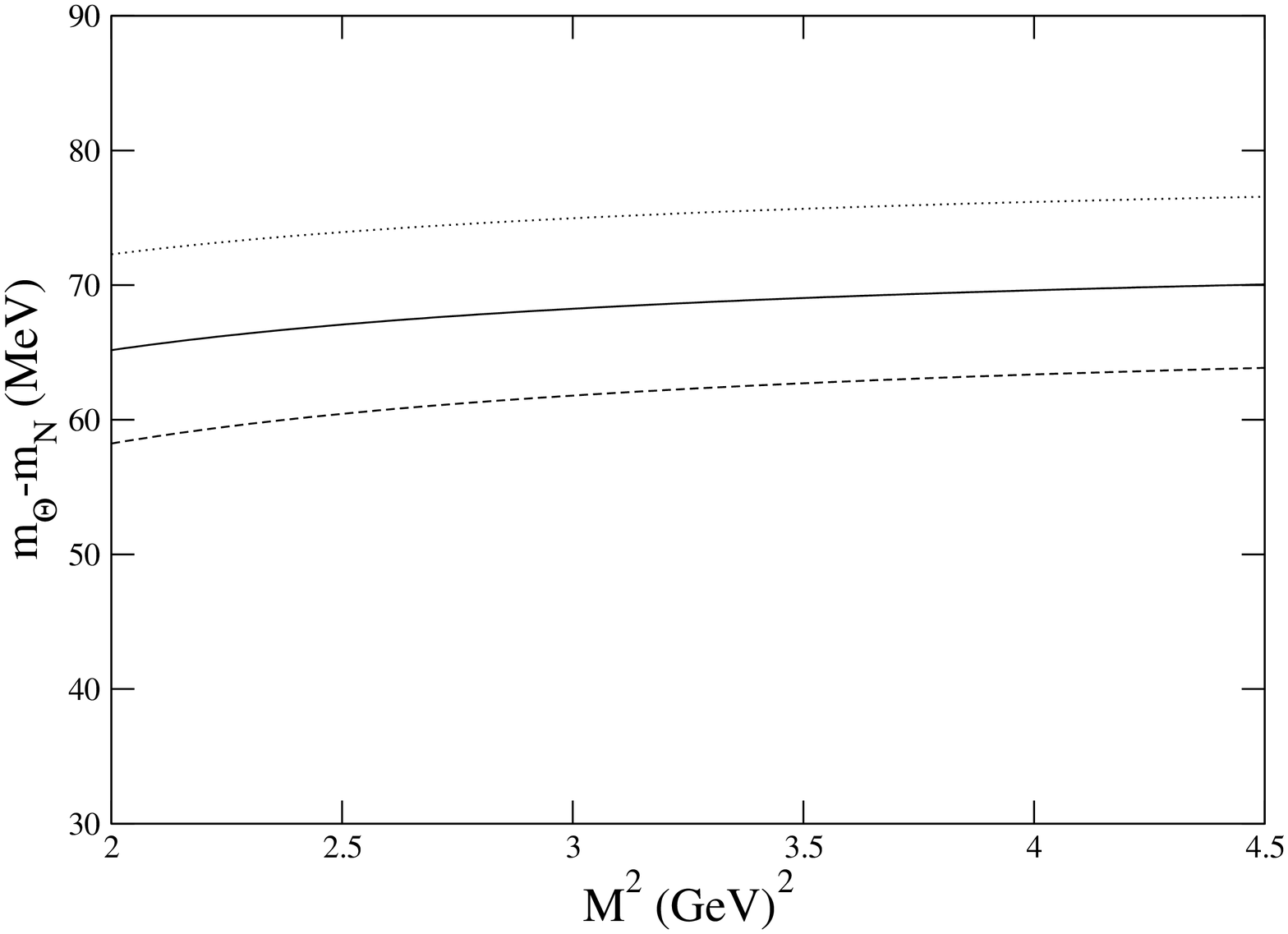}
\caption{\label{fig:3} Mass difference $m_\Theta-m_N$ for different
values of the continuum threshold, the solid, dashed and dotted lines
are for $s_{0\Theta}=3.5\ \gev^2$ and $s_{0N}=3.2\ \gev^2$,
$s_{0\Theta}=4.1\ \gev^2$ and $s_{0N}=3.8\ \gev^2$ and
$s_{0\Theta}=3.0\ \gev^2$ and $s_{0N}=2.7\ \gev^2$, respectively.}
\end{center}
\end{figure}
Similar as it has been done in \cite{Matheus:2003xr}, in fig.
\ref{fig:3} we plot the mass difference $m_\Theta-m_N$ for different values of
the continuum thresholds.
The mass splitting between the pentaquark states comes out to be about 70
MeV.
The error represented in fig. \ref{fig:3} is based on the assumption
that the continuum thresholds have the same offset for both pentaquark states.
Phenomenologically, these values can be different and one should add to the
error a part of the uncertainty from $s_0$ given in fig. \ref{fig:1}. 
Thus the error can easily amount to 50 MeV. Though the mass difference is 
consistent with the interpretation of the $N(1440)$ as a pentaquark,
the uncertainty remains large and a reduction of the error would be essential
to clarify the situation.

%%%%%%%%%%%%%%%%%%%%% Summary %%%%%%%%%%%%%%%%%%%%%%%%%%%%%%

To summarise, we have performed a QCD analysis based on the approach by Jaffe
and Wilczek. We obtain a sum rule that is stable over the Borel parameter
$M$ and reproduces the mass of the pentaquark within errors. 
We have also performed an analysis for the pentaquark with the quantum
numbers of the nucleon and have shown that the interpretation of the 
Roper resonance $N(1440)$ as $(ud)^2\bar{d}$ pentaquark state is 
consistent with the sum rules.
It is important to note that the sum rules are directly based on QCD and thus,
apart from the structure of the current, do not contain further model
assumptions. It would be interesting to see if lattice calculations could
confirm these findings. 
First lattice calculations exist \cite{Lattice} 
which, however, are based on
different interpolating currents and whose results are not yet conclusive.
A comparison and discussion can be found in \cite{Csikor:2004us}.
Hopefully, future experimental and theoretical investigations could further 
clarify and explore the pentaquark states and the consequences of the 
different models.

%%%%%%%%%%%%%%%%%%%%%%%%%%%%%%%%%%%%%%%%%%%%%%%%%%%%%%%%%%%%%%%%%%%%%%%%%%%%%%%

\section*{Acknowledgments}
I would like to thank S. Narison for the invitation to this
pleasant and interesting conference.
I thank the European Union for financial 
support under contract no. HPMF-CT-2001-01128. 
This work has been supported in part by 
EURIDICE, EC contract no. HPRN-CT-2002-00311 and by MCYT (Spain) under grant 
FPA2001-3031.

% \bibliography{ref_penta}

\end{document}